\documentclass[]{aa}
\usepackage[dvips]{color}
\usepackage{epsfig}
\usepackage{graphicx}

\def\deg {$^\circ$}

\titlerunning{ISGRI}
\authorrunning{Lebrun et al }

\begin{document}


\title{ISGRI: the INTEGRAL Soft Gamma-Ray Imager\thanks{Based on observations with INTEGRAL, an ESA project
with instruments and science data centre funded by ESA member
states (especially the PI countries: Denmark, France, Germany,
Italy, Switzerland, Spain), Czech Republic and Poland, and with
the participation of Russia and the USA. } }


\author{
 F. Lebrun\inst{1} \and
 J.P. Leray\inst{1} \and
 P. Lavocat\inst{1} \and
 J. Cr\'{e}tolle\inst{1} \and
 M. Arqu\`{e}s\inst{2} \and
 C. Blondel\inst{1} \and
 C. Bonnin\inst{1} \and
 A. Bou\`{e}re\inst{1} \and
 C. Cara\inst{1} \and
 T. Chaleil\inst{3} \and
 F. Daly\inst{1} \and
 F. Desages\inst{4} \and
 H. Dzitko\inst{1} \and
 B. Horeau\inst{1} \and
 P. Laurent\inst{1} \and
 O. Limousin\inst{1} \and
 F. Mathy\inst{2} \and
 V. Mauguen\inst{1} \and
 F. Meignier\inst{1} \and
 F. Molini\'{e}\inst{3} \and
 E. Poindron\inst{1} \and
 M. Rouger\inst{4} \and
 A. Sauvageon\inst{1} \and
 T. Tourrette\inst{1}
}

\date{Received date; Accepted date}
\offprints{F. Lebrun (flebrun@cea.fr)\\} \institute{
 CEA-Saclay, DSM/DAPNIA/Service d'Astrophysique, 91191 Gif-sur-Yvette Cedex, France, \and
 CEA-Grenoble, LETI, 17, rue des Martyrs, Grenoble Cedex 9, France, \and
 CEA-Saclay, DSM/DAPNIA/SIS, 91191 Gif-sur-Yvette Cedex, France, \and
 CEA-Saclay, DSM/DAPNIA/SEDI, 91191 Gif-sur-Yvette Cedex, France }


\abstract{For the first time in the history of high energy
astronomy, a large CdTe gamma-ray camera is operating in space.
ISGRI is the low-energy camera of the IBIS telescope on board the
INTEGRAL satellite. This paper details its design and its
in-flight behavior and performances. Having a sensitive area of
2621 cm$^2$ with a spatial resolution of 4.6 mm, a low threshold
around 12 keV and an energy resolution of $\sim$ 8\% at 60 keV,
ISGRI shows absolutely no signs of degradation after 9 months in
orbit. All aspects of its in-flight behavior and scientific
performance are fully nominal, and in particular the observed
background level confirms the expected sensitivity of 1 milliCrab
for a 10$^6$s observation.

\keywords{Space Telescope, Cadmium Telluride Detectors, Gamma-ray
Astronomy, Calibration, \textit{INTEGRAL}, IBIS.} } \maketitle


\section{Introduction}

A spectral coverage from several tens of keV to several MeV was
one of the main requirements for the INTEGRAL imager IBIS
(Ubertini et al., 2003). It is difficult with a single detector
and its electronic chain to cover more than two decades in energy.
For that reason, the IBIS detection unit uses two gamma cameras,
ISGRI covering the range from 15 keV to 1 MeV and PICsIT (Labanti
et al. 2003) covering the range from 170 keV to 10 MeV. This paper
describes the ISGRI gamma camera and reports its ground
performance and its flight behaviour. The in-flight calibration is
reported by Terrier et al. (2003). Detectors in space are affected
for some time after the passage of charged particles such as
cosmic-ray protons that deposit a huge amount of energy. A very
large detector such as the gamma camera of the SIGMA telescope on
board GRANAT (Paul et al. 1991) is crossed several times per
millisecond. As a result, the overall performances, and
particularly the spatial resolution, are degraded at low energy.
Pixel gamma-cameras, where each pixel is an independent detector
with its own electronic chain, avoid this problem since the
average time between two successive protons in a single detector
can be relatively long; allowing for a complete recovery of the
electronics. Moreover, the angular resolution of pixel gamma
cameras is independent of energy and can be made as good as
permitted by the power consumed and dissipated by the large number
of electronic chains. This and the need to ensure a low threshold
below 20 keV were the main drivers for the design of the ISGRI
gamma camera. The spectral performance, the ability to operate at
ambient temperature and the technological maturity of the cadmium
telluride (CdTe) manufacturing led to the choice of this
semi-conductor that was never used to build a large gamma camera
neither in space nor even on ground.

\begin{figure*}[t]
  \begin{center}
   \epsfig{file=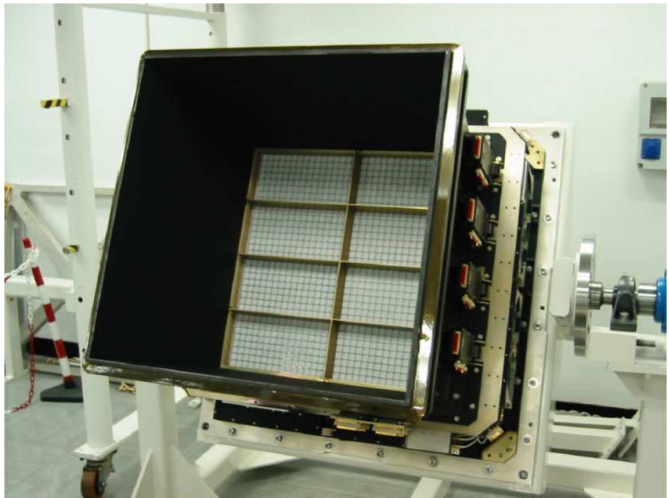,width=\columnwidth}
  \end{center}
    \vspace {-0.6cm}
 \caption{View of the 8 ISGRI MDUs (white color) at the bottom of
 the passive shield well (black color) after integration in the
 IBIS detection unit at LABEN premises. Four DBBs (black boxes with
 a red connector cap) are visible on the shield side (Courtesy IAS).}
  \label{pol. back}
\end{figure*}


\section{Instrument requirements}\label{sec:ISGRI req.}

The scientific requirements concerned the energy range, the
sensitivity, the spatial resolution, the timing accuracy and the
spectral performance. The energy range should offer an ample
overlap with the X-ray monitor JEM-X (Lund et al., 2003) on one
side and PICsIT on the other side. The sensitivity should be close
to that attained by OSSE (Johnson et al., 1993). The spatial
resolution should allow a sufficient sampling of the mask pattern
in order not to degrade the sensitivity. The timing accuracy
should be good enough to allow digital coincidences between ISGRI
and PICsIT. There were no strong requirements on the spectral
performance. As far as engineering requirements were concerned,
the total power consumption was limited to 130 W and the weight to
$\sim$ 30 kg. The experiment should be able to sustain the space
and launch conditions, i.e. temperature range, particle
irradiation, electromagnetic interferences and vibrations. An
important reliability requirement was that no single point failure
could induce a loss of data exceeding 20\%.


\section{Instrument design}

Figure 1 is a view of the detection plane of the ISGRI camera
formed with 8 independent modules. Each pixel of the camera is a
CdTe detector read out by a dedicated integrated electronic
channel. Altogether, there are 16384 detectors (128$\times$128)
and as many electronic channels. Each detector is a 2 mm thick
CdTe:Cl crystal of 4$\times$4 mm by side with platinum electrodes
deposited with an electroless (chemical) process. The ACRORAD
company provided 35 000 detectors in total for the various models
of ISGRI. All detectors have been screened for their spectroscopic
performance and stability under a 100 V bias at 20\deg C. Observed
instabilities led to the rejection of about 10\% of the detectors.
With regard to the spectral performance, the lot has been found
very homogeneous. Degradation of the spectroscopic performance of
the detectors under proton irradiation has been evaluated with
accelerator tests (Lebrun et al. 1996) and found not to be a
concern in the context of the INTEGRAL mission.

\begin{figure}[b]
  \begin{center}
    \epsfig{file=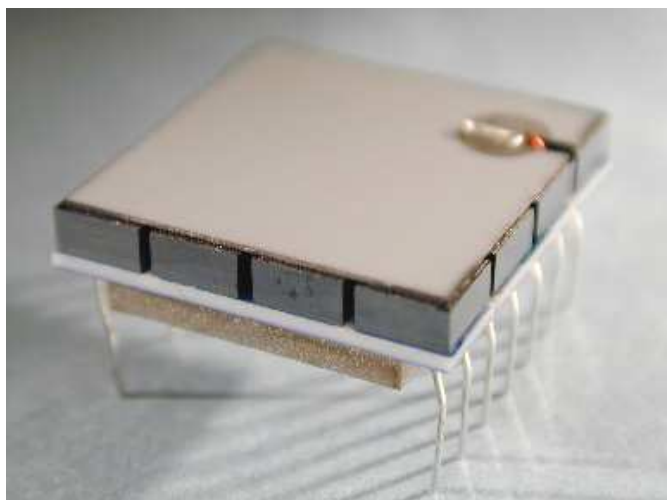,width=\columnwidth}
  \end{center}
    \vspace {-0.6cm}
 \caption{Front side of one of the 1024 ISGRI polycells. The 4$\times$4
 detectors can be seen in sandwich between the ceramic plates.}
  \label{pol. front}
\end{figure}

The large difference in the electron and hole mobilities in CdTe
implies that the pulse rise-time depends on the interaction depth.
As a result, a wide range in the pulse rise-time (0.5 -- 5 $\mu$s)
is observed and the adopted shaping time is a compromise between
the response to fast and slow pulses. Celestial low energy
photons, below 50 keV, always induce fast pulses. At higher
energy, slow pulses may be induced but are still less numerous
than the fast ones. As a result, a shaping time around 1.5 $\mu $s
has been chosen and the ballistic losses are important for slow
pulses. The optimum temperature for the detector stability and
spectral performance is around 0\deg C. PICsIT having a similar
requirement, IBIS and INTEGRAL have been designed to provide an
operational range around this value. In view of the number of
independent channels, integrated electronics are necessary. A low
noise and very low power consumption (2.8 mW/channel) Application
Specific Integrated Circuit (ASIC) has been designed for the
readout of 4 channels (Arques et al. 1999). It allows for the
simultaneous measurement of  the height and the rise time of every
pulse. The RMS noise of the preamplifier is ~165 e$^-$ (1pF input
capacitance) allowing good spectral performance down to $\sim$ 12
keV. This chip has many functionalities. For every channel, the
low-threshold (0-70 keV) and the gain are both adjustable as well
as the ASIC output gain and high threshold (around 1 MeV). An
event significantly above 1 MeV may saturate the preamplifier and
destabilize the baseline. For that reason, a system to reset the
ASIC and restore the baseline was implemented since cosmic-ray
particle may generate very high energy deposits. Internal
generators can simulate 60 keV and 600 keV events of various rise
times. Finally, the output of every channel can be disabled. The
layout used a radiation hardened library of components so that the
chip is latch-up free and only weakly sensitive to Single Event
Upsets (SEU). Sixteen detectors are mounted on one side of a
multilayer ceramic plate that hosts 4 ASICs on the other side. The
spacing between detectors is 600 $\mu$m. The detectors are
connected to the ASICs through vias in the ceramic substrate. The
electronics is protected with an hermetic titanium cover. An
alumina plate with a copper flash covers the detectors to provide
the bias ($\sim$ 100 V) and radiates away part of the heat
dissipated by the ASICs. The thickness of every component (except
detectors) have been minimized to optimize the transparency to
gamma-rays. As a result, this 4$\times$4 micro camera, called
polycell, shown in figure 2, weights less than 5g. The polycells
are glued and their twelve pins soldered on a multi-layer printed
circuit board stiffened with an aluminium grid. Each cell of the
grid contains a polycell. There are 8 rows of 16 polycells each. A
Field Programmable Gate Array (FPGA) allows to disable each of the
8 polycell-rows. With 2048 detectors, the so formed Modular
Detection Unit (MDU) represents 1/8 of the ISGRI detection area.
Two thermal probes are placed on the MDU frame, one on the long
side (TEMP1) and one on the short side (TEMP2).

Each MDU is connected to a Detector Bias Box (DBB) that delivers a
very stable bias voltage in the range 60 V -- 160 V and routes the
commands and signals to a Module Control Electronics (MCE). The
MCE configures the ASICs, encodes the pulse height and rise time
and processes the housekeeping data. Multiple triggers within a
single MDU are discarded (after encoding). The ASIC configuration
is stored in a context table containing the pixel status,
thresholds and gains as well as the MDU bias.

The detector stability cannot be guaranteed at 100\% and this was
a concern since a noisy detector triggering continuously precludes
the detection of gamma rays. A Noisy Pixel Handling System (NPHS)
was therefore implemented in the MCE to automatically switch off
(or raise the low threshold) of noisy pixels. This system has got
one counter per pixel (PC) and one counter per module (MC). Two
maximum values, PCM and MCM are defined. If MC reaches MCM, all
counters are reset but if one PC (out of 2048) reaches PCM, the
relevant pixel low threshold is raised by one step before the
counter reset, unless it is already at the maximum value (LTM). In
that case, the NPHS disables the relevant pixel and resets all
counters. Once per hour every pixel having a low threshold lower
than step 63 (70 keV) is enabled. If the pixel is noisy, it will
be disabled immediately, on the other hand, if quiet it will stay
ON. This allows for the recovery of pixels that are noisy only for
a short period. This dynamic tuning of each MDU guaranties the
best MDU functionality in spite of the unpredictable detector
behavior. The PCM and MCM values governs the NPHS sensitivity.
Values of 3 for the PCM and 100 for the MCM were perfectly adapted
to the detector behavior in the laboratory. Each MDU with its DBB
and MCE forms a fully independent gamma camera. This independence
implies that any single point failure cannot induce a loss of more
than 12.5\% of the sensitive area. The data relative to single
events from each MCE are transferred to the ISGRI Fifo Data
Manager (IFDM) that encodes the arrival time of the events, sorts
the events in time and transfers them to the IBIS Hardware Event
Processing Interface. This HEPI ensures the coincidence with
PICsIT, the high energy camera, and transfers the data to the Data
Processing Electronics (DPE) that formats the telemetry. Trigger
rates, temperatures and technological parameters are directly
transmitted to the DPE in the housekeeping data. The IFDM, HEPI
and DPE are cold redundant to maintain the overall reliability. To
reject the prompt background due to cosmic-ray protons, the IBIS
detection unit is surrounded with a VETO BGO detector that
delivers a programmable strobe to the IFDM. The ISGRI events in
coincidence with this strobe are discarded. A $^{22}$Na tagged
source provides an on-board calibration system. This calibration
source delivers also a strobe to the IFDM. Events in coincidence
with the calibration source are marked in the IFDM with a
calibration flag. The programmable delay and width of the VETO and
calibration strobes are identical. Four MCEs and one IFDM are
fitted in an ISGRI Electronics Box (IEB). There are two IEBs, all
MCEs being connected to the two IFDMs for redundancy. After
selection on the rise time (low and high thresholds) and the
energy (high threshold), all the information relative to every
valid ISGRI event (time, detector address, pulse height, pulse
rise-time) is downlinked.
\begin{figure*}[t]
  \begin{center}
    \hspace {-1.0cm}
   \epsfig{file=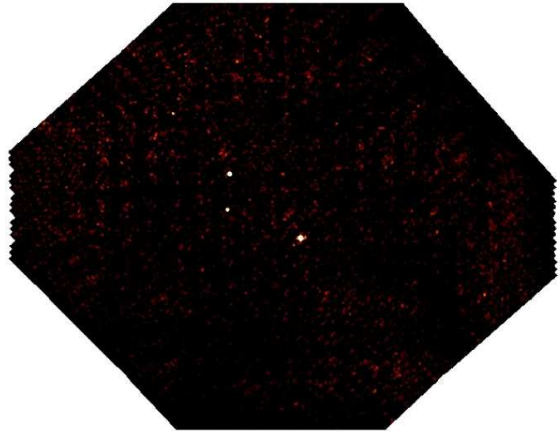,width=\columnwidth}
  \end{center}
    \vspace {-0.6cm}
 \caption{ISGRI image of the Cygnus region in the 15--40 keV energy range.}
  \label{bipar}
\end{figure*}

\section{Camera performance}\label{sec:Cam}

    \subsection{In flight detector behaviour}\label{sec:Cam_E}

Since launch, the measured temperature at the detector-layer level
is comprised between $-$10\deg C and +10\deg C, i.e. very close to
the expected temperature. The hole mobility is smaller at lower
temperature and consequently, the pulse rise-time increases. As a
result, ballistic losses are higher and the energy resolution is
worse. Increasing the bias voltage allows to maintain the pulse
rise-times in the filter bandpass. While in the laboratory at
20\deg C the bias was usually set to 100 V, it was set to 120 V in
flight. The first camera switch-on in orbit was done
progressively, MDU by MDU. When the first one was switch-on, it
was noticed that the NPHS was switching-OFF pixels at a very high
rate, nearly one per second, so that the MDU was shortly nearly
blind. This was the result of isolated bursts of triggers in
single detectors, appearing randomly in the MDU, hardly
attributable to detector behavior. This is most likely the result
of preamplifier overload may be due to $\sim$10 MeV electrons
passing through the mask holes. The maximum number of consecutive
triggers observed in a detector was about one hundred. Setting the
PCM to 200 and the MCM to 10 000 allowed a full recovery of the
MDU functionality.

However, with these values, the NPHS is much less sensitive and
the pixel low thresholds are less accurately adjusted. A fine
adjustment of the low thresholds is performed at the ISDC on the
basis of the pixel spectra registered in the previous orbit
(Terrier et al. 2003). For the few tens of detector per MDU
exhibiting unreasonable spectra, the low thresholds were set to
step 63, i.e. disabled. After an initial period with a fast
increase of the number of disabled pixels, some sort of saturation
was reached with $\sim$ 400 detectors disabled. The pulse
rise-time of the consecutive triggers is always zero. To avoid a
telemetry overload, the on-board software was modified to reject
events with a pulse rise-time lower than a predefined value
(rise-time low threshold). After 9 months in orbit, the ISGRI
detectors show no sign of degradation. The only significant change
with regard to the ground behavior is the thermal gradient along
the detection plane. This is due to vacuum conditions that change
the thermal balance (Terrier et al., 2003). This proves that CdTe
can be safely used in space.

    \subsection{Imaging performance}\label{sec:Im. perf.}

  The sensitivity of a coded mask telescope depends on the ability
  of the camera to finely sample the shadowgram. This so called imaging
  efficiency is governed by the ratio of the smallest mask elements and
  the detector spatial resolution. For pixel cameras such as ISGRI
  and PICsIT it is energy independent and should be insensitive to the
  space conditions. This is a definitive advantage over the Anger
  gamma camera where a monolithic scintillator is read out by many photomultiplier tubes.
  As a matter of fact, the Anger gamma camera spatial resolution varies with energy and is strongly
  degraded by the effect of cosmic-ray protons.
The 4 mm ISGRI pixels are spaced every 4.6 mm ensuring, with 11.2
mm mask elements, an excellent imaging efficiency of 0.86. With a
mask-ISGRI distance of 3.2 m the angular resolution is $12'$ and a
10$\sigma$ source can be localized with an accuracy of $1'$. The
total ISGRI area is around 3600 cm$^2$ but dead zones between
pixels and between MDUs restrict the sensitive area to 2621
cm$^2$. In addition, about 3\% of the pixels are switched off at
any given time, increasing accordingly the dead zone area. Finally
some pixels have a threshold higher than the minimum. Below their
threshold, they can be considered as disabled. The data
processing, in particular the image deconvolution takes all these
effects into account (Goldwurm et al. 2003). As a result, ISGRI
produces the finest images obtained so far in the soft gamma-ray
domain. This performance is illustrated in figure 3 that shows a
picture of the Cygnus region in the energy range 15-40 keV where
at least 3 sources are clearly visible and studied in other papers
of this volume: Cygnus X-1 (Bazzano et al. 2003), Cygnus X-3
(Goldoni et al. 2003) and EXO2030-338 (Kuznetsov et al. 2003).

    \subsection{Timing accuracy}\label{sec:Timing perf.}
ISGRI timing is used to establish coincidences with the PICsIT,
VETO and calibration source and also to provide useful scientific
information. ISGRI records the event time with a 240 ns precision.
However, there is a very important jitter, mainly due to the
variety in pulse rise-times, resulting in a FWHM timing accuracy
of the order of 2.5 $\mu$s. The strobe width has been adjusted to
5 $\mu$s as a compromise between the VETO efficiency and the
induced dead time. Due to telemetry limitations the relative
accuracy accessible for source timing studies is degraded to 61
$\mu$s. Taking into account systematic errors the final absolute
timing accuracy is of the order of 90 $\mu$s (at 3$\sigma$).

    \subsection{Energy range}\label{sec:Spect. perf.}
The low energy threshold of IBIS/ISGRI results from the
detector/electronics noise and the transparency of the intervening
material (mask, polycell caps). As far as the camera is concerned,
a low threshold of 12 keV is attained. However the opacity of the
mask is very low at this energy and the low threshold was
conservatively set to 13 keV. Moreover, the noise renders the data
processing more difficult close to the low threshold. The precise
value of the effective low threshold remains to be determined but
is certainly lower than 15 keV (the Crab is clearly seen between
14 and 17 keV). The high threshold allowed by the electronics is
$\sim$ 1 MeV and was never changed.

    \subsection{Spectral response}\label{sec:Spect. perf.}

 In a pixel camera such as ISGRI, every pixel is a spectrometer
chain with its own characteristics. The spectral performance of
the camera depends therefore critically on the alignment of the
pixel gains and offsets. This alignment is performed in two steps.
First, the electronics allow for a rough alignment and second, a
fine software correction must be applied. This and the temperature
effect is detailed in the next section. Since Compton scattering
plays a large role, the spectral response of a gamma-ray
instrument operating in the soft gamma-ray domain is always
complex. The spectral response of IBIS and ISGRI in particular was
estimated with Monte-Carlo simulations carefully built to
reproduce the calibration measurements (Laurent et al., 2003).

    \subsubsection{ISGRI MDU tuning and ground calibration}\label{sec:Cam_Ecal}

The 8 flight and the spare MDU have been successively mounted in 3
test benches ensuring a thermal control between $-$15\deg C and
30\deg C. First the ASICs gains were adjusted to minimize the
dispersion in the pulse-height and rise-time gains. The ASIC
configuration was then stored in the ISGRI context that is loaded
in the ASICs every 4 s to minimize the effect of a SEU in flight.
The actual pulse height and rise-time gains and offsets were
determined for every pixel to allow fine software corrections that
are mandatory for a proper spectral response of the whole MDU.
Then the average gains of each MDU were set by resistor adjustment
and the MDU calibration could proceed. Since the flight
temperature was not precisely known, spectral calibration
measurements were performed using various radioactive sources at
different temperatures and bias voltages. This represents hundreds
of measurements and more than 50 000 hours of data acquisition.
The dependence of the gains and offsets with temperature and bias
voltage was carefully calibrated. While the offsets are almost
independent of the temperature, the pulse-height and rise-time
gain variations are respectively 0.4\% deg$^{-1}$ and 3.3\%
deg$^{-1}$. Gain corrections for temperature variations are
therefore necessary and are based on the MDU thermal probe
measurements.

    \subsubsection{Charge loss correction}\label{sec:Cam_Ecal}

\begin{figure}[t]
  \begin{center}
    \hspace {-1.0cm}
   \epsfig{file=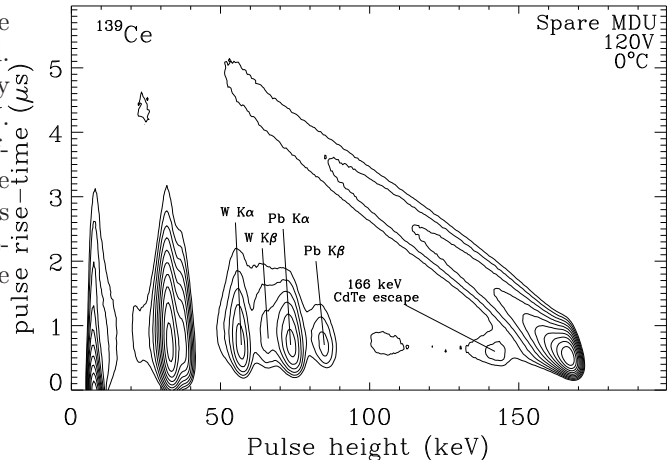,width=\columnwidth}
  \end{center}
    \vspace {-0.6cm}
 \caption{Contour plot of the count per pulse-height and pulse rise-time bins.}
  \label{bipar}
\end{figure}

Because of the charge loss, the ISGRI spectral response is rather
unusual above 60 keV. This is illustrated in figure 5 that
displays a pulse-height spectrum obtained with a collimated
$^{139}$Ce source. The line at 166 keV exhibits a plateau in its
left wing at half the height of the line. On the other hand, the
33 keV blend of lines doesn't have this wide left wing. This
high-energy behavior is due to the ballistic losses that are
important for slow pulses, i.e. due to interactions close to the
anode. The longer is the rise-time, the more important are the
losses. At 33 keV, all interactions take place close to the
cathode and produce fast pulses. The charge loss is negligible in
this case. At 166 keV, a significant fraction of the interactions
occurs near the anode and the pulse height alone can no longer
represent the energy deposited. Measuring simultaneously the pulse
height and the pulse rise-time allows for a proper energy
estimate. The relationship between the charge loss and the pulse
rise-time is illustrated in figure 4 that displays the measured
pulse height as a function of the pulse rise-time. Below 60 keV,
lines appears as vertical ellipsoids in this bi-parametric
diagram. Their vertical elongation is due to the uncertainty in
the pulse rise-time measurement that is more difficult for weak
pulses. At higher energy, lines appear as inclined tracks. From
this, it can be seen that a 166 keV energy deposit occuring close
to the anode (rise time $\sim$ 5 $\mu$s) gives a pulse of 50--60
keV. These bi-parametric diagrams are the basis of the charge loss
correction. Each point of this diagram, a couple of pulse-height
and pulse rise-time, should correspond uniquely to an energy. This
is generally the case except in the top-left region of the diagram
where medium and low energy can be mixed. The effect of this
correction is illustrated with the spectrum displayed in figure 6.
Rejecting events with rise-time greater than 4.5 $\mu$s avoids
most of the confusion region above mentioned and represents only a
loss of a few percent in efficiency. By comparison with figure 5,
one notes that the plateau below the 166 keV line is strongly
reduced and that the peak is significantly enhanced. Athough the
peak is much more symmetric, large wings are present in the line
profile. They are due to long rise-time events that have a much
worse energy resolution. The line shape is not gaussian and the
FWHM should be taken with care. Selecting only events with
rise-time shorter than 0.6 $\mu$s improves drastically the
spectral performance (see Fig. 7) at the expense of the efficiency
that is reduced by $\sim$ 70\%. In this case, the line profile is
almost gaussian.

\begin{figure}[t]
  \begin{center}
    \hspace {-0.4cm}
    \epsfig{file=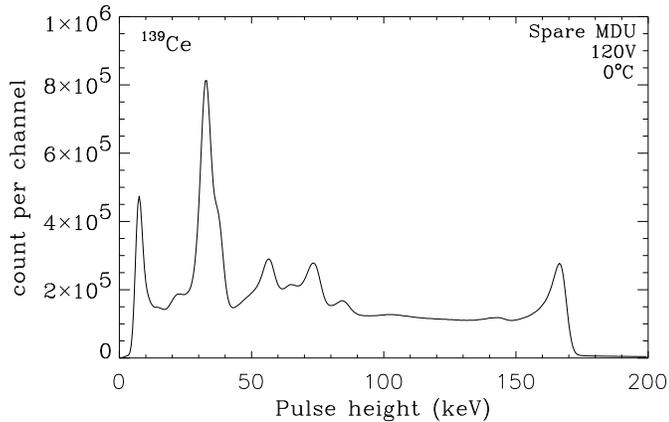,width=\columnwidth}
  \end{center}
    \vspace {-0.6cm}
 \caption{Pulse height spectrum acquired at 0\deg C with the spare MDU biased at 120 V and illuminated by a $^{139}$Ce source.}
  \label{raw sp}
\end{figure}

\begin{figure}[t]
  \begin{center}
    \epsfig{file=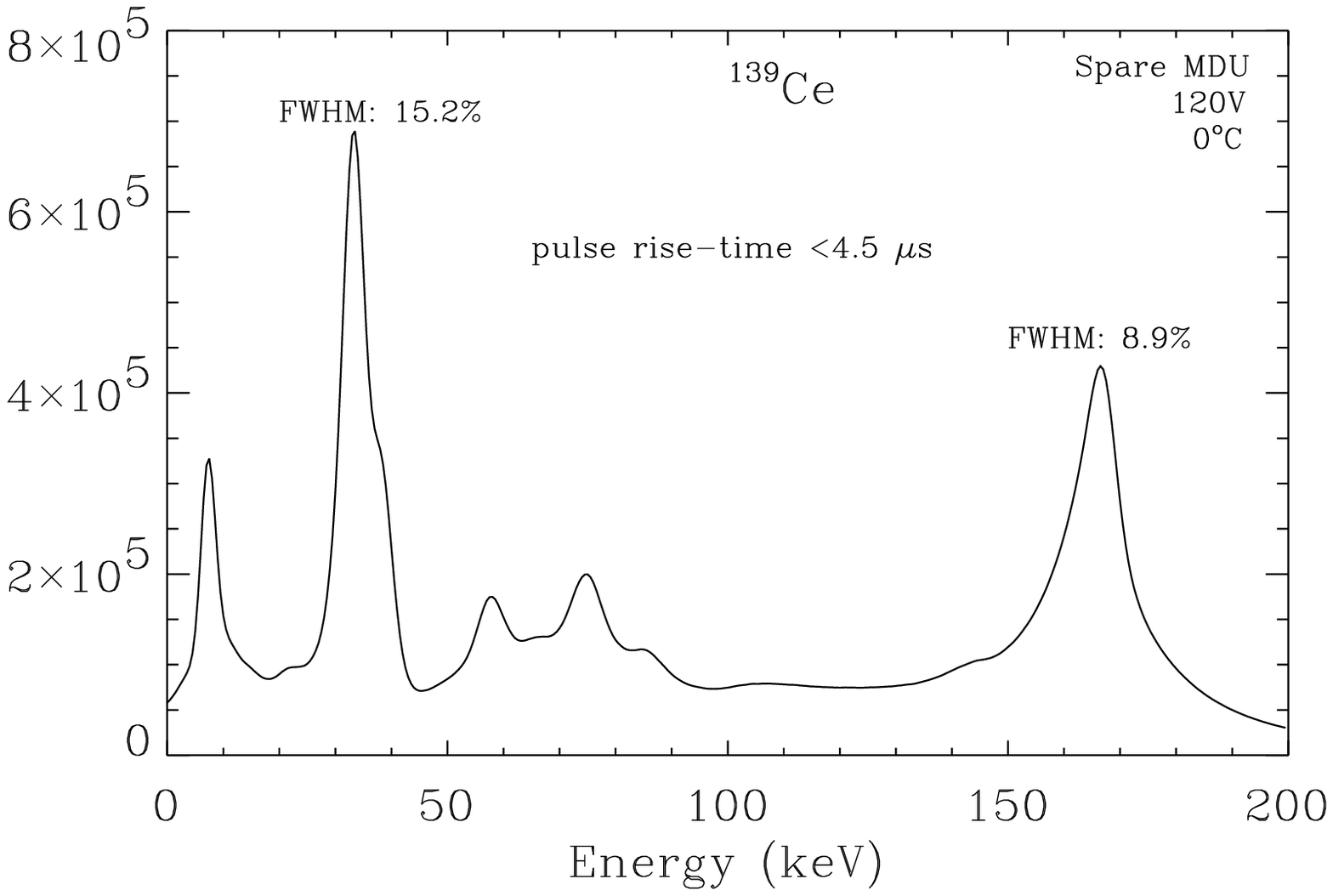,width=\columnwidth}
  \end{center}
    \vspace {-0.6cm}
 \caption{$^{139}$Ce spectrum corrected for the charge loss. Events with pulse rise-time longer than 4.6 $\mu$s have been discarded.}
  \label{sp. corr.}
\end{figure}

\begin{figure}[t]
  \begin{center}
    \epsfig{file=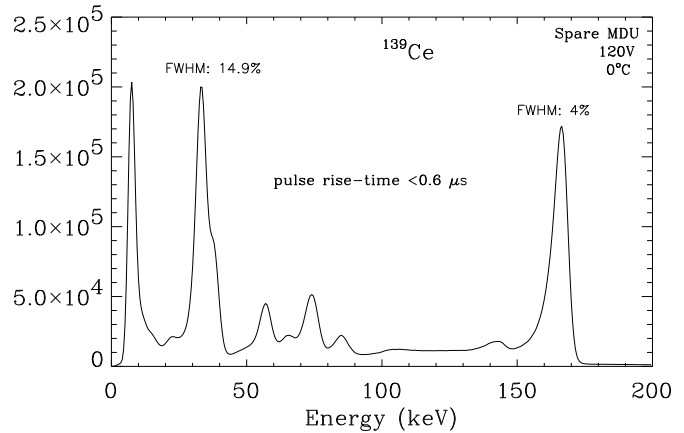,width=\columnwidth}
  \end{center}
    \vspace {-0.6cm}
    \caption{$^{139}$Ce spectrum corrected for the charge loss. Events with pulse rise-time longer than 0.6 $\mu$s have been discarded.}
  \label{sp. corr-sel}
\end{figure}

    \subsubsection{Spectral performance}\label{sec:Cam_Ecal}

The spectral performance depends on the temperature and the bias
voltage. Measurements on the flight model MDUs have been performed
at various temperatures but mostly with a 100 V bias. Figure 8
illustrates the ISGRI spectral resolution as a function of energy.
This is based on measurements performed on MDU 2 with a 100 V bias
at 0\deg C. Since the performance depends on the selection applied
on the pulse rise-time, two extreme cases have been considered: no
selection and only pulse rise-time equal to 0.5$\mu$s. While there
is almost no difference up to 60 keV, the spectral performance
with no selection degrades rapidly above this energy to reach a
maximum difference of a factor of 3 near 250 keV. Then the
spectral performance for short pulses is almost independent of
energy with $\Delta$E (FWHM)/E around 0.03. Measurements at 0\deg
C with various bias have been performed using a radioactive source
of $^{139}$Ce both on MDU 2 and on the spare MDU. The results are
compatible indicating a 7\% improvement in the energy resolution
(without rise-time selection) when rising the bias from 100 V to
120 V. There is a small degradation with the time after the bias
is set ($\sim$ 4\% after 100h). Possible drifts in the detector
response can be tracked thanks to the on-board tagged radioactive
source of $^{22}$Na that produces photons at 511 and 1275 keV and
induces lines at 60 and 75 keV from the fluorescence of tungsten
and lead (Terrier et al. 2003).
\begin{figure}[t]
  \begin{center}
    \epsfig{file=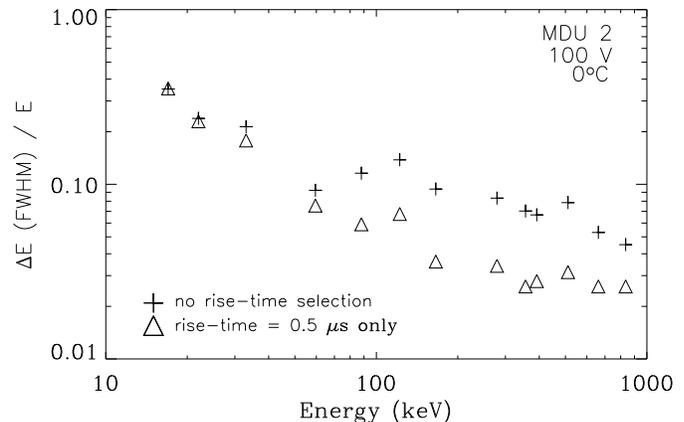,width=\columnwidth}
  \end{center}
    \vspace {-0.6cm}
    \caption{Spectral performance of ISGRI estimated from MDU 2.
    Two extreme cases are considered: without any selection on
    the rise time and with rise time equal to 0.5 $\mu$s only.}
  \label{sp. corr-sel}
\end{figure}

    \subsection{ISGRI Sensitivity}

    \subsubsection{Efficiency}\label{sec:Effic.}

The efficiency of the ISGRI camera depends first on the detector
efficiency. Although very thin, 2 mm, the detection efficiency of
the ISGRI detectors is still around 50\% at 150 keV. This is due
to the high atomic numbers of Cd and Te and also to the high
density, $\sim$ 6 g cm$^{-3}$, of the CdTe. However, the overall
efficiency depends also on the intervening absorbing material, the
mask and the polycell caps. Although every effort has been given
to maximize their transparency, they have a major effect at low
energy. The overall efficiency is given in figure 9. Due to the
opaque mask elements, the overall efficiency is reduced by a
factor of 2. The imaging efficiency was also taken into account in
the overall efficiency. Below 20 keV, the mask transparency
becomes so weak that the effective low threshold of ISGRI is not
anymore governed by the detector and electronic noise.
\begin{figure}[t]
  \begin{center}
    \epsfig{file=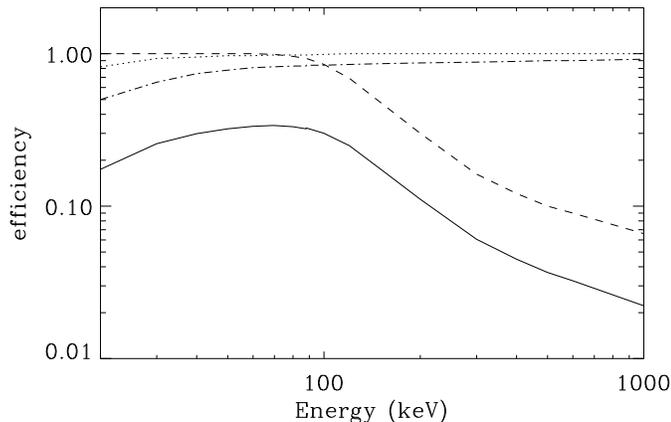,width=\columnwidth}
  \end{center}
    \vspace {-0.6cm}
    \caption{ISGRI detection efficiency (solid line). The detector
    efficiency (dashed line), the mask hole transparency
    (dashed-dotted line) and the polycell cap transparency
    (dotted line) are also given. Note that since half of the
    mask elements are opaque, the ISGRI detection efficiency
    has been divided by a factor two  }
  \label{sp. corr-sel}
\end{figure}

   \subsubsection{Dead time}\label{sec:Dead time}

The overall ISGRI dead time is due to the ISGRI event processing
time and to the coincidence applied with PICsIT, VETO and the
calibration source. The optimum anticoincidence width, depending
on the background level, has been determined in flight during the
INTEGRAL commissioning phase to be 5$\mu$s .The ISGRI event
processing time is 114 $\mu$s and the trigger rate per module is
around 800 s$^{-1}$ so that the associated dead time is around
9\%. Even more significant is the dead time due to the
coincidences. The VETO and the calibration source count rates are
respectively $2.5 \times 10^4$ s$^{-1}$ and $5.6 \times 10^3$
s$^{-1}$ inducing dead times of 12.5\% and 2.8\% respectively. The
PICsIT count-rate is $4.5 \times 10^3$ s$^{-1}$, with a
coincidence width of 3.8$\mu$s it induces a dead time on ISGRI of
1.7\%. The VETO, CAL and PICsIT events being independent, the
overall ISGRI dead time is of the order of 24\%.

    \subsubsection{Background}\label{sec:BKG}
In the case of ISGRI, the most critical and uncertain performance
parameter was the background spectrum. The low energy part
(E$<$100 keV) is dominated by extragalactic emission and is
relatively well known (Kinzer et al., 1997; Watanabe et al. 1997).
On the other hand, the high energy part of the spectrum is
dominated by the internal background. It is due mainly to the
de-excitation of nuclei produced by the spallation reactions of
cosmic-rays on the instrument. It was uncertain because it could
not be extrapolated from a previous mission and Monte-Carlo
simulations have not yet reached the required accuracy. It was
critical not only because it impacts directly on the experiment
sensitivity but also because it could induce a telemetry overflow.
This spectrum displayed in figure 10 is close to the expectations.
\begin{figure}[t]
  \begin{center}
    \epsfig{file=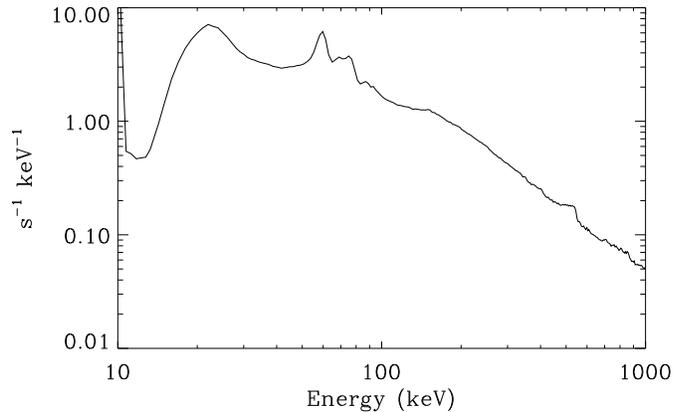,width=\columnwidth}
  \end{center}
    \vspace {-0.6cm}
    \caption{ISGRI background spectrum observed while INTEGRAL was pointing to an empty field.}
  \label{sp. corr-sel}
\end{figure}

    \subsubsection{Broad-band sensitivity}\label{sec:BKG}

The ISGRI sensitivity can be computed from the observed background
spectrum (Fig. 10) and the above reported efficiency (Fig. 9) and
dead time estimate. Figure 11 gives this ISGRI broad-band
($\Delta$E=E/2) sensitivity at 3$\sigma$ for an observing time of
$10^6$s.

\begin{figure}[t]
  \begin{center}
    \epsfig{file=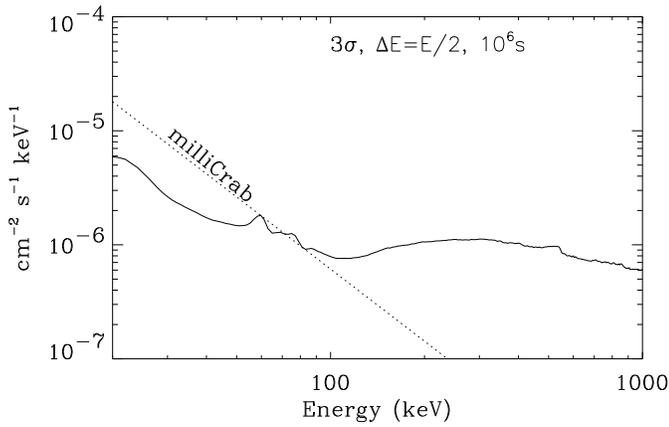,width=\columnwidth}
  \end{center}
    \vspace {-0.6cm}
    \caption{ISGRI broad-band sensitivity.}
  \label{sp. corr-sel}
\end{figure}


\section{Conclusion}

Today, after 9 months of in-flight operations, there are no signs
of detector degradation. The ISGRI performance is fully nominal.
The observed background, very close to the expectations, implies a
milliCrab sensitivity for a $10^6$s observing time. CdTe was known
for its very good potential as a gamma-ray spectrometer and it is
now proven that a large detector can be realized and safely used
in space. The ISGRI camera produces the best images ever obtained
in the soft gamma-ray domain. In the coming years, CdTe (or
CdZnTe) will undoubtedly play a key role in instrumental
high-energy astrophysics.


\begin{acknowledgements}

The authors would like to thank
\begin{itemize}
    \item CNES
    \item the IAS team in Roma
    \item the LABEN team in Milano
    \item and the INTEGRAL project team at ESA
\end{itemize}
 for their indefectible support
during the difficult phases of the programme.%
\end{acknowledgements}




\end{document}